\documentclass[twocolumn,prl,superscriptaddress,showpacs]{revtex4}

\usepackage{graphicx}
\usepackage{mathrsfs}
\usepackage{amssymb}
\usepackage{color}

\newcommand{\psin}{\rho}

\begin{document}

\title{Transition dynamics in aging systems: microscopic origin of
logarithmic time evolution}

\author{Michael A. Lomholt}
\affiliation{MEMPHYS, Department of Physics, Chemistry and Pharmacy,
University of Southern Denmark, DK-5230 Odense M, Denmark}
\author{Ludvig Lizana}
\affiliation{Department of Physics and Center for Soft Matter Research,
New York University, 4 Washington Place, New York, NY 10003, USA}
\affiliation{Integrated Science Lab, Department of Physics, Ume{\aa}
University, SE-901 87 Ume{\aa}, Sweden}
\author{Ralf Metzler}
\affiliation{Institute for Physics \& Astronomy, University of Potsdam,
D-14476 Potsdam-Golm, Germany}
\affiliation{Department of Physics, Tampere University of Technology, FI-33101
Tampere, Finland}
\author{Tobias Ambj{\"o}rnsson}
\affiliation{Department of Astronomy and Theoretical Physics, Lund
University, SE-22362 Lund, Sweden}

\begin{abstract}
There exists compelling experimental evidence in numerous systems for
logarithmically slow time evolution, yet its theoretical understanding
remains elusive. We here introduce and
study a generic transition process in complex systems, based
on non-renewal, aging waiting times. Each state $n$ of the system follows a
local clock initiated at $t=0$. The random time $\tau$ between clock ticks
follows the waiting time density $\psi(\tau)$. Transitions between states
occur only at local clock ticks and are hence triggered by the local forward
waiting time, rather than by $\psi(\tau)$. For power-law forms $\psi(\tau)\simeq
\tau^{-1-\alpha}$ ($0<\alpha<1$) we obtain a logarithmic time evolution of the
state number $\langle n(t)\rangle\simeq\log(t/t_0)$, while for $\alpha>2$ the
process becomes normal in the sense that $\langle n(t)\rangle\simeq t$. In the
intermediate range $1<\alpha<2$ we find the power-law growth $\langle n(t)
\rangle\simeq t^{\alpha-1}$. Our model provides a universal description for
transition dynamics between aging and non-aging states.
\end{abstract}


\date{\today}

\maketitle

Imagine that you put a thin sheet of paper in a vertical cylinder and let the
paper crumple under a heavy piston. If during compression you measure the
piston's velocity you will notice that it decreases over time, well in
accordance with your intuition. However, what may appear surprising is that
the piston keeps compressing the paper and never seems to come to a full rest.
The outcome of such an experiment was reported by Matan
et al.~\cite{matan2001crumpling}, concluding that the piston's position $z(t)$
at long times $t$ decreases logarithmically, $z(t)\sim a-b\log(t/{\sec})$,
where $a$ and $b$ are constants. The crumpling of paper is by far the only
example for logarithmically slow dynamics. It is observed in DNA local structure
relaxation \cite{brauns2002complex}, the time evolution of frictional strength
\cite{ben2010slip}, compactification of grains by tapping
\cite{richard2005slow}, kinetics of amorphous-amorphous transformations in
glasses under high pressure \cite{tsiok1998logarithmic}, magnetisation
dynamics in high-$T_c$ superconductors \cite{gurevich1993time}, conductance
relaxations \cite{amir2012relaxations, amir2011huge} and current relaxation in
semiconductor field-effect transistors \cite{woltjer1993time}. Other examples
of logarithmic time evolution include decays in colloidal systems
\cite{sperl2003}, aging in simple glasses \cite{angelani01} (see also
Supplementary Material \cite{supp}), magnetisation relaxation in spin glasses
\cite{chowdhury1984logarithmic}, evolution of node connectivity in a
network with uniform attachment \cite{barabasi1999}, diffusion in a random
force landscape (Sinai diffusion) \cite{havlin2002}, and record statistics
\cite{zia1999}.

\begin{figure}
\includegraphics[width=8cm]{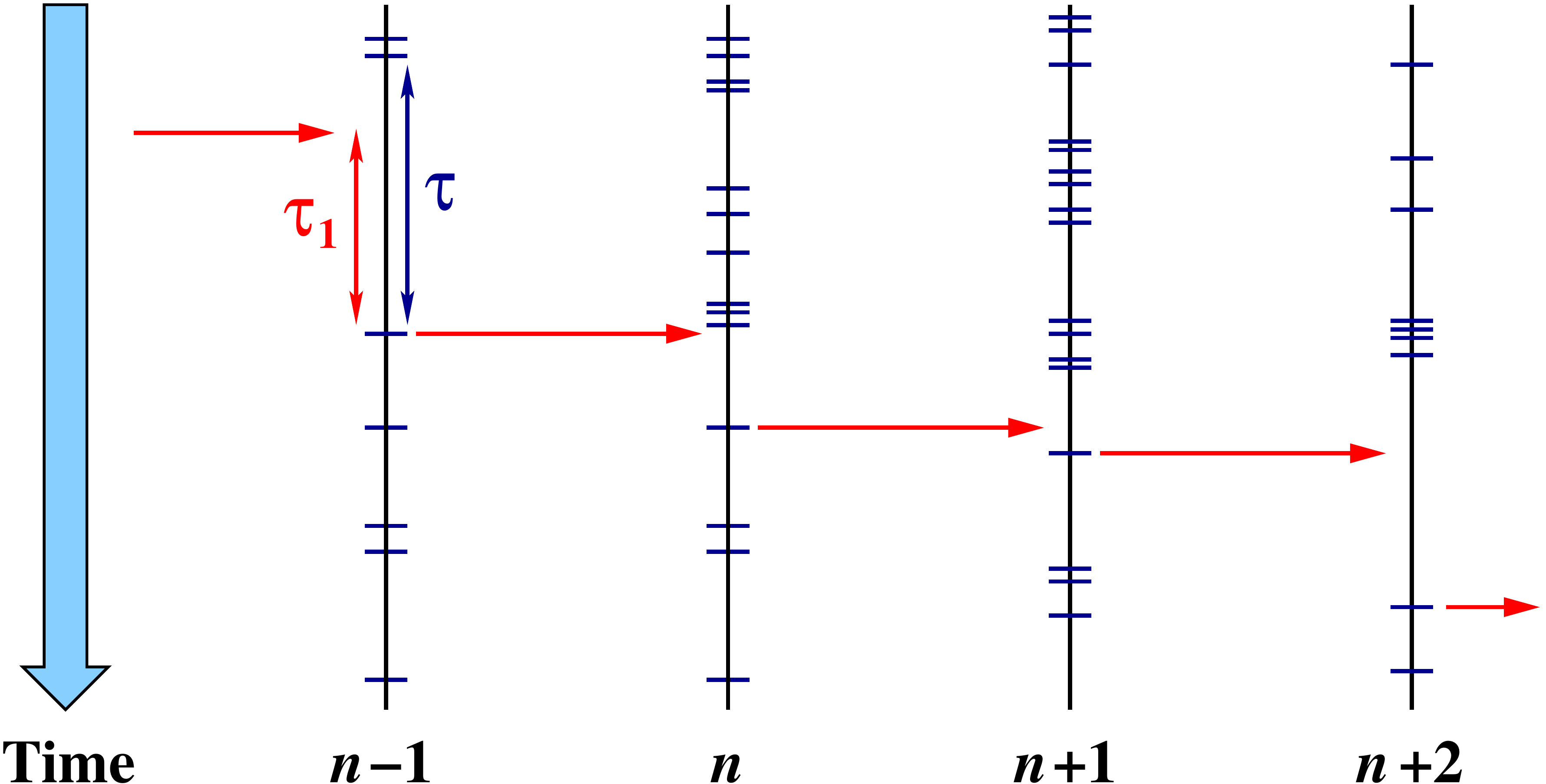}
\caption{Dynamic update of successive states. At each state $n$, a tick of the
local clock allows the transition to the next state, $n+1$. Local clock ticks
are separated by waiting times $\tau$ drawn from the distribution $\psi(\tau)$.
After transition from state $n-1$ the system is locked in $n$ until the next
clock tick at $n$, after the forward waiting time $\tau_1<\tau$. Typically a
transition at a new state arrives during a long waiting time, the statistics of
the $\tau_1$ thus slowing down the overall dynamics.}
\label{fig1}
\end{figure}

Here we introduce a generic microscopic model displaying logarithmic time
evolution which is based on non-renewal sequential transitions between aging
states labeled ny $n$ (see Fig.~\ref{fig1}). As we show analytically, the $q$th order
moments of the resulting counting process at large times grow as
\begin{equation}
\label{eq:mainres}
\langle n^q(t)\rangle\sim\left[\frac{\ln(t/t_0)}{\mu}\right]^q\left\{1+
\frac{q}{2}\left[\frac{q\sigma^2}{\mu}-\mu\right]\frac{1}{\ln(t/t_0)}\right\}
\end{equation}
such that, particularly, we find the logarithmically slow counting process
$\langle n(t)\rangle\simeq\log(t/t_0)$. The parameters $\mu$ and $\sigma$
depend on the details of the underlying dynamics and are specified below,
and $t_0$ is the time when the counting started after global system initiation at $t=0$, for instance, by an
external perturbation.
We also show that under non-aging conditions our model leads to the expected
linear growth $\langle n(t)\rangle\simeq t$, and in the intermediate case we
observe power-law scaling for $\langle n(t)\rangle$.
Our model provides
an intuitive mesoscopic approach to the superslow dynamics in aging systems.

We define the dynamics of the system through a series $n(t)$ of consecutive
states, each of which is characterized by its own local clock and all being
initiated globally at time $t=0$. The clocks' ticks occur with random time
intervals $\tau$, which are drawn from a waiting time density
$\psi(\tau)$ (see Fig.~\ref{fig1}). If the system arrives at state $n-1$ at
a later time $t'$ then it is more likely to
a encounter a large $\tau$, and therefore also typically has to wait a
correspondingly longer time $\tau_1$ before a transition to state $n$ occurs.
For $\psi(\tau)\simeq\tau^{-1-\alpha}$
with $0<\alpha<1$, no typical time scale $\langle\tau\rangle=\int_0^{\infty}
\tau\psi(\tau)d\tau$ exists, and we find Eq.~(\ref{eq:mainres}), whose scaling with
the counting initiation time $t_0$ manifests the aging property of the
process \cite{REMM}.
Equation~(\ref{eq:mainres}) is the central result of this work, but
we also obtain $\langle n^q(t)\rangle$ for $\alpha>1$. Moreover, we find the
probability distribution $h_n(t)$ to be in state $n$ at time $t$ given that
the counting of transitions (from state $0$) began at $t=t_0$.

A simplistic picture for our model is to envision a hitchhiker traveling
through a series of towns. In each town, traffic starts in the morning,
and friendly drivers (persons willing to pick up our hitchhiker) appear
at random intervals $\tau$ governed by $\psi$. The hitchhiker typically
arrives to a new town in between two friendly drivers show up, and the
delay time $\tau_1$, i.e. the time the hitchhiker actually has to wait
until the next ride, is governed by the forward waiting time density $\psi_1$
\cite{REM}. The probability density $\psi_1$ is far from trivial: for
heavy-tailed $\psi(\tau)$ it displays aging, see below. In this context it is
interesting to note that indeed arrival times of English trains, but also
response times in human communication patterns, and bursting in queuing models
are power-law distributed \cite{REMMM,oliveira2005human,barabasi2005origin}.

A more physical picture for our model is defect-mediated crack-type propagation in
a solid. Imagine a crack that grows in discrete steps ($...,n-1,n,...$), the
growth being triggered by the arrival of a diffusing defect at the
neighbouring site of the crack's tip, similar in spirit to Glarum's defect
diffusion model \cite{glarum}. The global initiation in this system occurs
when the external stress is applied. Possibly, similar scenarios may apply
in the above-mentioned examples of stick-slip dynamics \cite{ben2010slip} and
density relaxation of grains by tapping \cite{richard2005slow}.

We now formulate our process mathematically. To that end we define the probability
density $\psin_n(t)$ for the system to arrive at state $n$ at time $t$, which
fulfils the convolution
\begin{equation}
\label{eq1}
\psin_n(t)=\int_0^t \psin_{n-1}(t')\psi_1(t-t'|t')d t',\,\,
\psin_0(t)=\delta(t-t_0),
\end{equation}
where $\psi_1(\tau_1|t')$ is the probability density of the triggering delay
time (forward waiting time) $\tau_1$ that the system spends in a new state
after having arrived there at time $t'$. Equation (\ref{eq1}) expresses the
fact that the probability to arrive at state $n$ in a time interval $[t,t+dt]$
is the probability of having arrived to the state $n-1$ at some earlier time
interval $[t',t'+dt']$ ($t'<t$) multiplied by the probability of a triggering
event occurring in $[t',t'+dt']$, where $t'$ lies anywhere between 0 and $t$.
Now, if $\psi(\tau)\simeq\tau^{-1-\alpha}$ with $0<\alpha<1$ ($\alpha>1$ is
discussed below) then the
probability density $\psi_1$ of forward waiting times $\tau_1$ is known from
continuous time random walk (CTRW) theory, namely \cite{dynkin,godreche,koren}
\begin{equation}
\label{eq:leapoverdist}
\psi_1(\tau_1|t')=\frac{\sin(\pi\alpha)}{\pi}\frac{{t'}^\alpha}{\tau_1^{
\alpha}(t'+\tau_1)}.
\end{equation}
This quantity explicitly depends on the arrival time $t'$ and thus mirrors the
aging property of the process: while at small $t'$, we observe the scaling
$\psi_1\simeq\tau_1^{-1-\alpha}$ in analogy to the regular waiting time density
$\psi(\tau)$, at longer $t'$ we have to wait for a longer $\tau_1$ for the 
next transition event. This intuitively corresponds to the observation of a
random walk process governed by the waiting time density $\psi(\tau)\simeq
\tau^{-1-\alpha}$ with $0<\alpha<1$: when the process evolves (i.e., becomes
older), due to the scale-free nature of $\psi$ we see increasingly longer
waiting times. The later we arrive at a new state (growing $t'$), the longer 
will the current tick-tick waiting time $\tau$ be and thus $\tau_1$ grows
longer as the overall process develops.

We note that our model is in stark contrast with standard CTRW theory where
the waiting time is reset (renewed) after each transition \cite{report,scher},
i.e., the renewals are an intrinsic property of the process. Here we update
each state \emph{locally\/} starting at $t=0$, and each local clock is renewed
after a tick. However, the overall process effectively couples all the local
clocks, since after a transition to a new state $n$ (i.e., a tick at state
$n-1$) the process needs to wait for the next local tick (at $n$). This
bestows the non-renewal property of the overall process.

Finally, we obtain the probability $h_n(t)$ to find the system in state $n$ at
time $t$. It corresponds to the probability of having arrived at $n$ at $t'<t$,
and no transition having occurred since:
\begin{equation}
\label{eq:hnt}
h_n(t)=\int_0^t\psin_n(t')\int_{t-t'}^\infty\psi_1(\tau_1|t')d\tau_1 d t'.
\end{equation}
Eqs.~(\ref{eq1}) to
(\ref{eq:hnt}) define the problem we solve here. 

To proceed it is convenient to employ the technique of Mellin transforms
\cite{davies2002}. With $G(x)\equiv x\psi_1(x-1|1)\theta(x-1)$, where
$\theta(x)$ is the unit step function, Eq.~(\ref{eq1}) becomes
\begin{equation}
\label{eq:rewrit}
\psin_n(t)=\frac{1}{t} \int_0^\infty\psin_{n-1}(t') G(t/t') d t'.
\end{equation}
Using the definition of Mellin transforms $f(p)=\int_0^\infty t^{p-1}f(t)dt$,
where $p$ is the Mellin variable, along with the Mellin convolution theorem
\cite{davies2002} we obtain from Eq.~(\ref{eq:rewrit}) that $\psin_n(p)=G(p-1)
\psin_{n-1}(p)$, to which the solution is $\psin_n(p)=[G(p-1)]^n t_0^{p-1}$
[here we used $\psin_0(p)=t_0^{p-1}$]. The Mellin transform of
Eq.~(\ref{eq:hnt}) is $h_n(p)=\psin_n(p+1)[G(p)-1] /p$, and therefore
\begin{equation}
\label{eq:h_n}
h_n(p)=t_0^pG(p)^n[G(p)-1]/p.
\end{equation}
This is an exact solution in Mellin space for the sought-after quantity $h_n(t)$
used in the following.

While no simple expression exists for the exact $h_n(t)$ we can obtain
all moments of $h_n(t)$ in the limit of long times $t$. Expanding $G(p)$ for
small $p$ to second order, for $0<\alpha<1$, we obtain the $q$th order
moments \cite{supp}
\begin{equation}
\label{res}
\langle n^q(p)\rangle\sim \frac{\Gamma(q+1)t_0^p}{\mu^q(-p)^{q+1}}\left\{1+
\frac{p}{2}\left[\mu-q\sigma^2/\mu\right]\right\},\,\,\, p\to 0^{-}
\end{equation}
in Mellin space,
with $\mu=-\Gamma'(\alpha)/\Gamma(\alpha)-\gamma$ and $\sigma^2=-\pi^2/6+
\partial^2\ln\Gamma(\alpha)/\partial\alpha^2$. Here, $\Gamma(z)$ is the complete
$\Gamma$ function, and $\gamma=0.5772..$ denotes Euler's constant.
Inverting the Mellin transform, we retrieve Eq.~(\ref{eq:mainres}) at long $t$.
Thus the leading order behavior of the first two moments follows
$\langle n(t)\rangle\sim\ln(t/ t_0)/\mu$ and $\langle
n^2(t)\rangle\sim\ln^2(t/t_0)/\mu^2$. This shows that the triggering process
considered here leads to a non-trivial logarithmic time evolution for
heavy-tailed forms of $\psi(\tau)$. The logarithmically slow dynamics
contrasts the case $\alpha>1$ for which $\langle n^q(t)\rangle$ grows
as a power-law (shown below). In Fig.~\ref{fig2} we compare our analytical
result (\ref{res}) for $\langle n(t)\rangle$ with simulations \cite{REM1} for
the concrete form $\psi(\tau)=\alpha\tau_0^\alpha/(\tau+ \tau_0)^{1+\alpha}$.
As can be seen, the simulations agree excellently with Eq.~(\ref{eq:mainres}),
except for $\alpha\to1$. The inset of Fig.~\ref{fig2} shows that the mismatch
is due to the fact that $t_0$ is not sufficiently large (i.e., not much larger
than $\tau_0$) and the distribution $\psi_1(\tau_1|t_0)$ thus has not reached
its asymptotic form~(\ref{eq:leapoverdist}).

\begin{figure}
\begin{center}
\includegraphics[width=8cm]{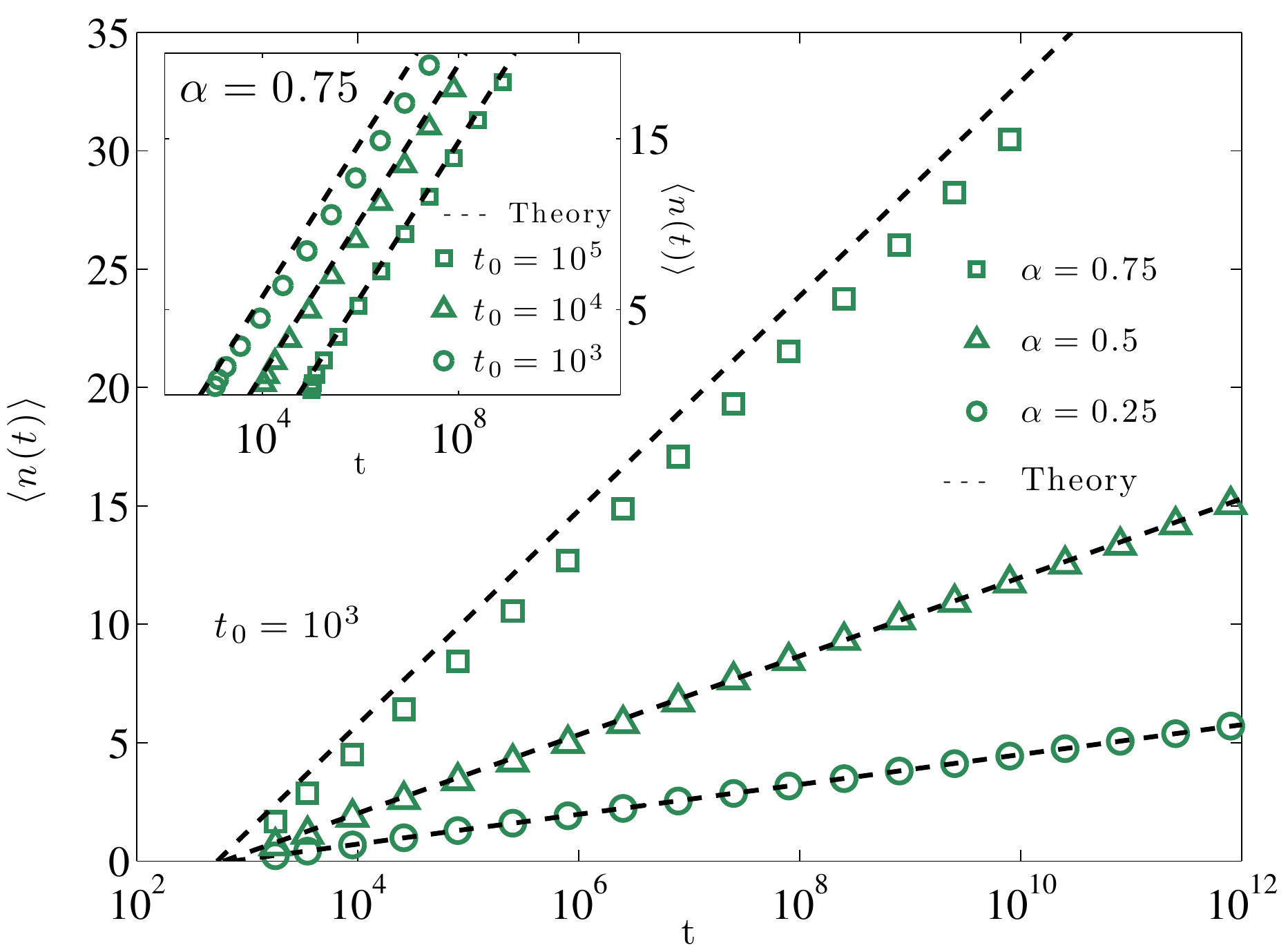}
\end{center}
\caption{Average state number $\langle n(t)\rangle$ versus time $t$. Symbols
represent simulations for various $\alpha$, as indicated. Dashed lines show
the asymptotic behavior, Eq.~(\ref{eq:mainres}) for $q=1$. In the simulations
we used $t_0=10^3$ and $\tau_0=1$. Results are ensemble averaged over $10^7$
runs, respectively. Inset: convergence to the theoretical results with $t_0$
for $\alpha=0.75$.
\label{fig2}}
\end{figure}

\begin{figure}
\begin{center}
\includegraphics[width=8cm]{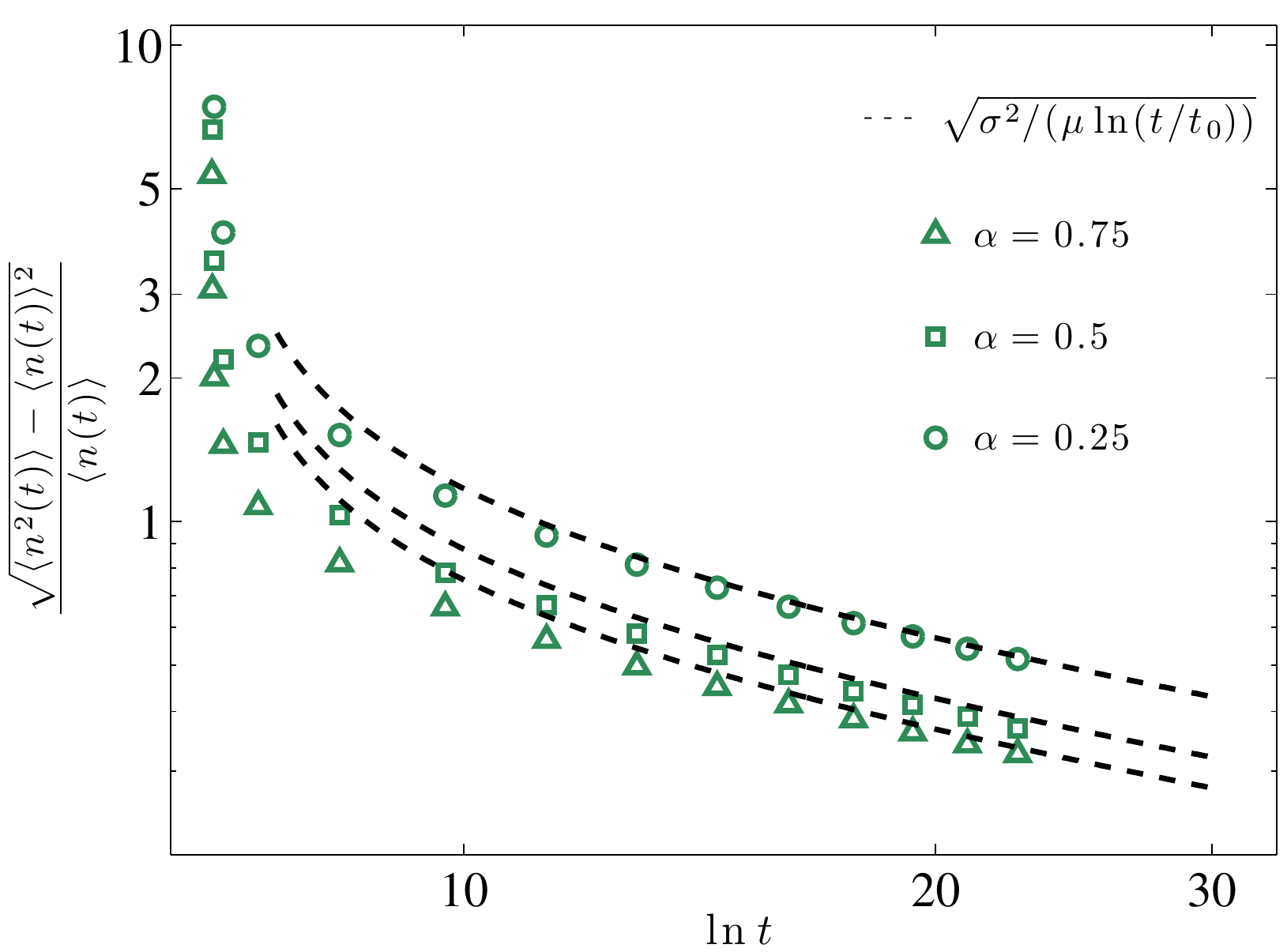}
\end{center}
\caption{Standard deviation versus mean as a function of process time.
Simulations (symbols) are compared to the asymptotic results for large
$t$ [Eq.~(\ref{rel_width}), dashed lines]. Parameters used: $t_0=10^3$,
$\tau_0=1$, averaged over $7\times 10^5$ simulation runs.
\label{fig4}}
\end{figure}

The $q$-dependence of the dominant term in Eq.~(\ref{eq:mainres}) corresponds
to a $\delta$-function for the limiting distribution. This means that the
standard deviation versus the mean in our model becomes increasingly small for
long times and that the dynamics becomes effectively \emph{deterministic}.
Indeed, dividing the variance by the mean we find 
\begin{equation}
\label{rel_width}
\frac{\sqrt{\langle n^2(t)\rangle-\langle n(t)\rangle^2}}{\langle n(t)\rangle}
\sim\sqrt{\frac{\sigma^2}{\mu\ln(t/t_0)}},
\end{equation}
as is nicely corroborated by simulations of this ratio in Fig.~\ref{fig4}.
Equation (\ref{rel_width}) contrasts the behavior of the position coordinate
in biased subdiffusive CTRW processes
where the ratio
above tends to a constant \cite{scher}.

What about the behavior when $\alpha>1$? In this case $\psi_1(\tau_1|t')$ has
a finite limit independent of $t'$ and is given by $\psi_1(\tau_1)=\int_{\tau_1}
^\infty\psi(\tau')d\tau'/\langle\tau\rangle$ \cite{klaso}, where $\langle\tau
\rangle=\int_0^{\infty}\tau\psi(\tau)d\tau$. Assuming the form $\psi(\tau)\sim
A/\tau^{\alpha+1}$ for large $\tau$ one obtains $\psi_1(\tau_1)\sim(\alpha-1)A/
[\langle\tau\rangle\tau_1^\alpha]$. We find two distinct regimes for the
cases $1<\alpha<2$ and $\alpha >2$. For $1<\alpha<2$ the system goes through
the series of states, $n(t)$, as a regular renewal process with power-law
waiting times of index $\alpha-1$. The number of states the system passes in
this case thus has the moments \cite{koren,REM2}
\begin{equation}
\label{eq:sublinear}
\langle n^q(t)\rangle\sim\frac{\Gamma(q+1)}{\Gamma(q(\alpha-1)+1)}\left(\frac{
\langle\tau\rangle t^{\alpha-1}}{\Gamma(2-\alpha)A}\right)^q \propto t^{
(\alpha-1)q}.
\end{equation}
Here we notice that the mean $\langle n(t)\rangle\simeq t^{\alpha-1}$ increases
sublinearly rather than logarithmically as in the case $0<\alpha<1$.
Moreover, we find that the fluctuations grow as fast as
the mean. For $\alpha>2$ we put $\alpha\to2$ and $\Gamma(2-\alpha)A\to\langle
\tau^2\rangle/2$ so that we obtain
\begin{equation}
\label{linear}
\langle n^q(t)\rangle\sim\left(\frac{2\langle\tau\rangle t}{\langle\tau^2
\rangle}\right)^q \propto t^q.
\end{equation}
In this case, in particular, the mean grows linearly with time. Interestingly,
just as for the case $0<\alpha<1$ (but in contrast to the regime $1<\alpha<2$),
the deviations vanish relative to the mean, i.e., the long-time dynamics is
effectively deterministic.

\begin{figure}
\begin{center}
\includegraphics[width=8cm]{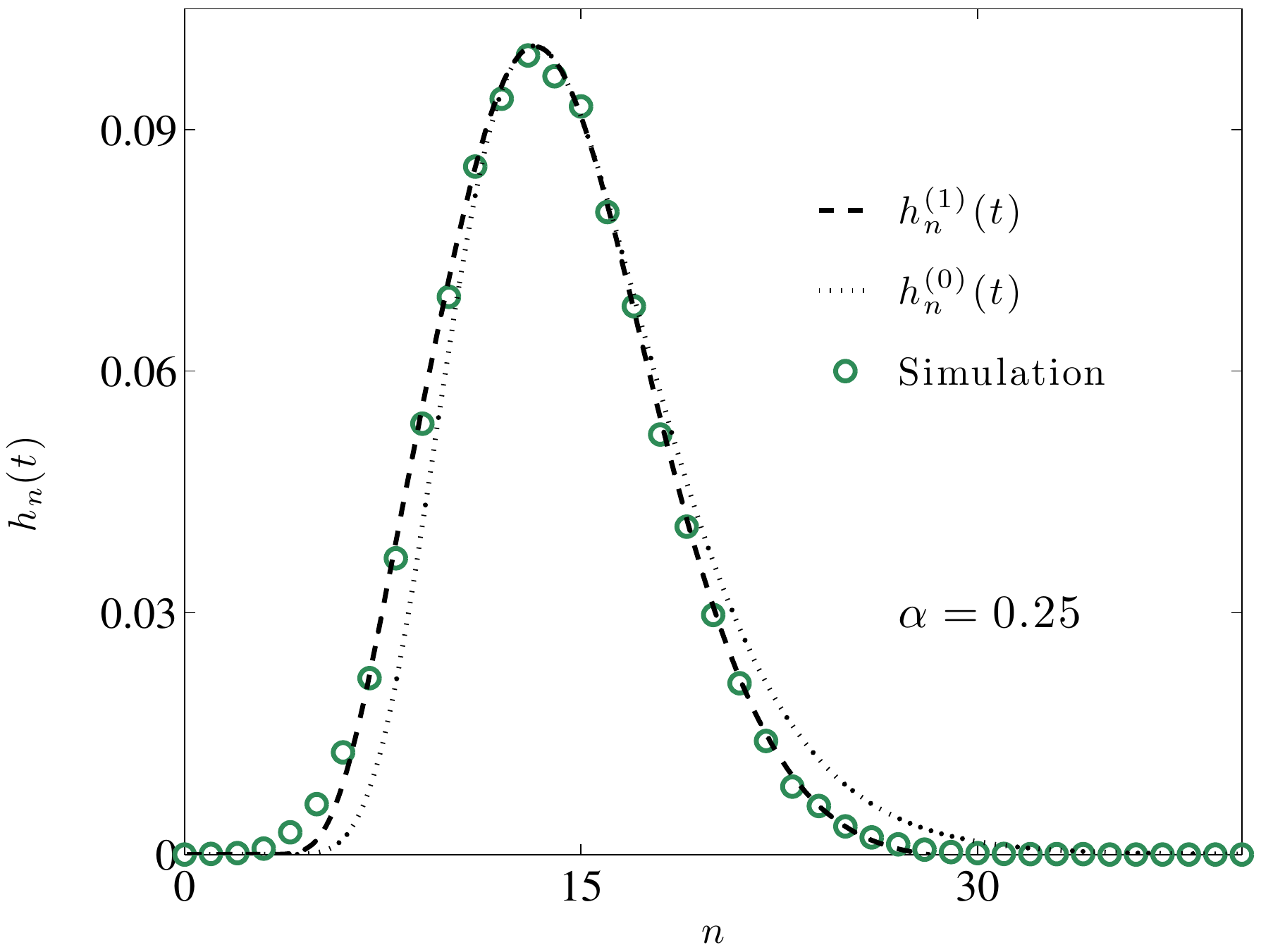}
\end{center}
\caption{Probability distribution to find the system in state $n$ at time $t$.
Lines: analytical results from Eqs.~(\ref{hnexp}) and (\ref{hn0}). Circles:
simulation results, averaged over $7\times 10^5$ runs. Parameters used: $t=7.9
\times 10^{12}$, $t_0=10^3$, $\tau_0=1$, and $\alpha=0.25$.
\label{fig5}}
\end{figure}

We now turn our attention to the full distribution $h_n(t)$ for the case
$0<\alpha<1$. To that end we need to evaluate the inverse Mellin transform of
Eq.~(\ref{eq:h_n}). In the Supplementary Material \cite{supp} we derive the
approximate form 
\begin{equation}
h_n^{(1)}(t)=h_n^{(0)}(t)\left[1+\frac{\sigma^2+\mu^2}{2\mu\sqrt{\sigma^2
n}}y+\frac{\kappa_3n}{6(\sigma^2n)^{3/2}}(y^3-3y)\right],
\label{hnexp}
\end{equation}
where $y=[\ln(t/t_0)-\mu n]/\sqrt{\sigma^2 n}$ and
\begin{equation}
h^{(0)}_n(t)=\frac{\mu}{\sqrt{2\pi\sigma^2n}}\exp\left(-\frac{(\ln(t/t_0)-\mu
n)^2}{2\sigma^2 n}\right).
\label{hn0}
\end{equation}
The distribution $h_n(t)$, for fixed (logarithmic) time, is thus a slightly
skewed Gaussian in the $n$-domain. In Fig.~\ref{fig5} we compare the result
$h_n^{(1)}(t)$ with simulations, demonstrating good agreement for its
dominating part.

In particle tracking assays single trajectories are routinely measured and
analysed \cite{pt}. We therefore also consider the time average for a single
realization of $n(t)$ defined as $\overline{n(\Delta)}=(t_2-\Delta-t_1)^{-1}
\int_{t_1}^{t_2-\Delta} [n(t+\Delta)-n(t)]dt$, where the observation time of
the trajectory is from $t_1$ to $t_2$, and $\Delta$ is the lag time. Here we
only consider the heavy-tailed case $0<\alpha<1$. Averaging over many
trajectories, the dominant behavior at $t_1,t_2\gg t_0$ becomes
\begin{equation}
\left<\overline{n(\Delta)}\right>\sim\Delta\frac{1/\mu}{t_2-t_1}\ln\frac{
t_2}{t_1}
\end{equation}
for $\Delta\ll t_2-t_1$.
The linear behavior in $\Delta$ contrasts the logarithmic time dependence of
$\langle n(t)\rangle$. 
This discrepancy between ensemble and time average demonstrates that the process considered
here is weakly non-ergodic \cite{pt,bouchaud92,bel07}. Interestingly, while the duration
$t_2-t_1$ and the aging time $t_1$ factorize from the lag time
($\Delta$) dependence similar to CTRW processes \cite{js}, the times $t_1$ and $t_2$
enter in terms of the non-trivial combination $(\ln t_2-\ln t_1)/(t_2-t_1)$.

We finally ask whether we can understand the logarithmic time evolution for
$\langle n(t)\rangle$. We show in the Supplementary Material \cite{supp} that
Eq.~(\ref{eq:rewrit}) can, after minor modifications, be interpreted as the
probability density for products of independent random variables. The
logarithmic time evolution follows from the fact that the product of many
random numbers approaches the log-normal distribution. Our work therefore
connects to the large number of scientific fields where this
distribution appears, see the review \cite{limpert2001}.

In summary, we developed a generic stochastic framework for systems exhibiting
logarithmic time evolution. Our system is initiated globally by some external
perturbation (stress, incipient light etc.) but transitions occur by updates
of \emph{local\/} clocks. Each transition to the following state is thus timed
according to the first waiting time. Consequently, the resulting process is
`\emph{super-aging\/}' in the sense that at each step a local aging period
passes. As a result we obtain a logarithmic time evolution for power-law
forms of the clock-update distribution $\psi$.

Examples of logarithmically slow dynamics are found in biological, mechanical
and electrical systems. No universal framework has yet been put forward for
such dynamics and only classes of systems have been identified where their
theoretical descriptions have little in common. They are based on, for
example, macroscopic phenomenological assumptions of the system's behaviour,
extreme value statistics or specific types of particle-particle
interactions. In this work we explore a new class, a generic transition
process between aging states, where the logarithmic dynamics is an emergent
property. We solved the problem exactly and showed results for the temporal
distribution to reach a state $n$ as well as the moments $\langle
n^q(t)\rangle$. Due to the generic yet simple nature of our model we are
confident that it will be applied in many scientific fields.

LL acknowledges the Knut and Alice Wallenberg (KAW) foundation for financial
support. TA are grateful to KAW and the Swedish Research Council. RM
acknowledges funding from the Academy of Finland (FiDiPro scheme).

\newpage
\noindent
\includegraphics[page=1,trim=1.91cm 1cm 1cm 1.5cm,clip]{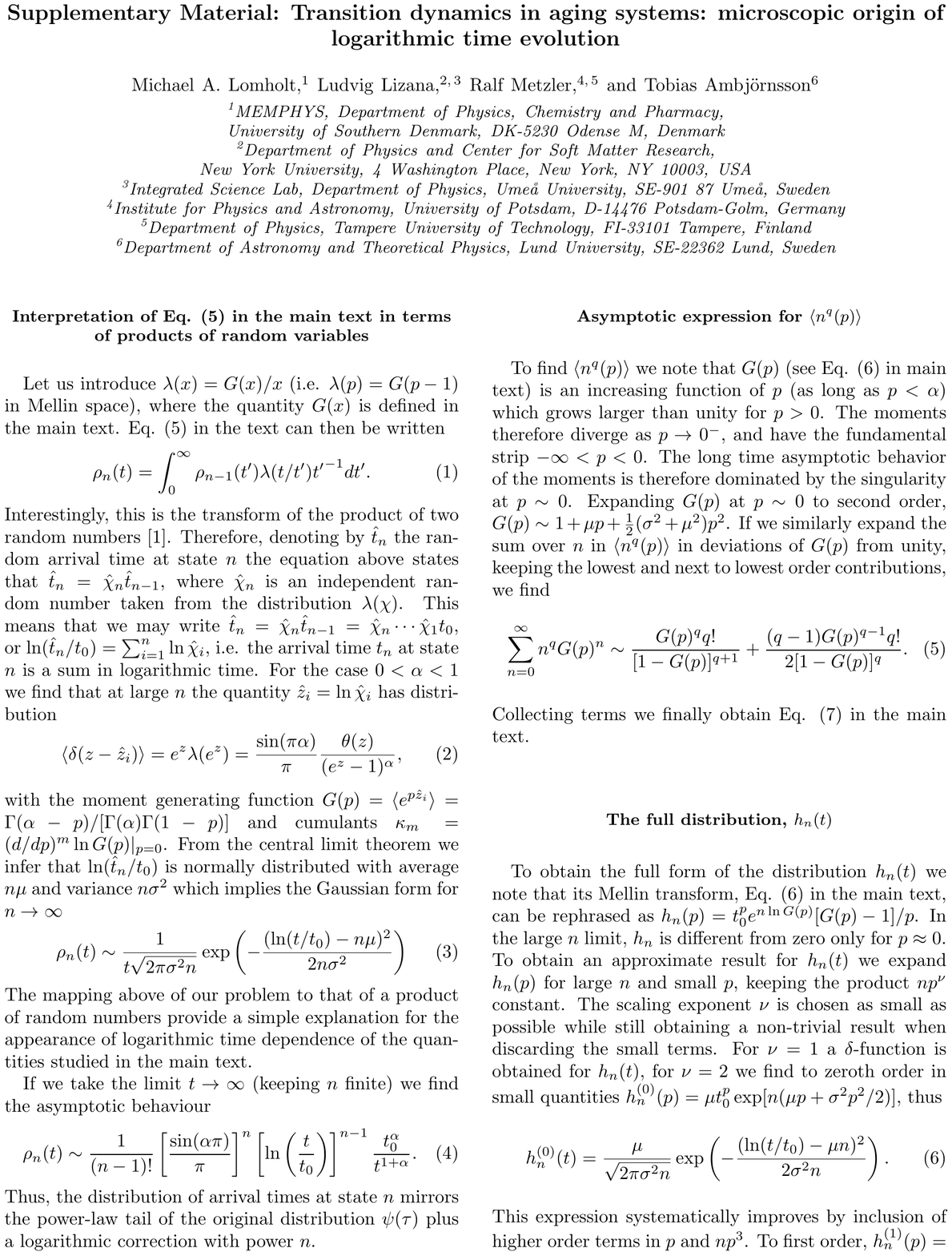}
\thispagestyle{empty}

\newpage
\noindent
\includegraphics[page=2,trim=1.91cm 1cm 1cm 1.5cm,clip]{supplementary.pdf}
\thispagestyle{empty}






\begin{thebibliography}{99}

\bibitem{matan2001crumpling} K. Matan, R. B. Williams, T. A. Witten,
and S. R. Nagel, Phys. Rev. Lett. {\bf 88}, 076101 (2002).

\bibitem{brauns2002complex} E. B. Brauns, M. L. Madaras, R. S. Coleman, C. J.
Murphy, and M. A. Berg, Phys. Rev. Lett. {\bf 88}, 158101  (2002).

\bibitem{ben2010slip} O. Ben-David, S. M. Rubinstein, and J. Fineberg,
Nature {\bf 463}, 76 (2010).

\bibitem{richard2005slow} P. Richard, M. Nicodemi, R. Delannay, P Ribi{\`e}re,
and D. Bideau, Nature Mat. {\bf 4}, 121 (2005).

\bibitem{tsiok1998logarithmic} O. B. Tsiok, V. V Brazhkin, A. G. Lyapin, and L.
G. Khvostantsev, Phys. Rev. Lett. {\bf 80}, 999 (1998).

\bibitem{gurevich1993time} A. Gurevich and H. K{\"u}pfer, Phys. Rev B.
{\bf 48}, 6477 (1993).

\bibitem{amir2012relaxations} A. Amir, Y. Oreg, and Y. Imry, Proc. Nat. Acad.
Sci. USA {\bf 109}, 1850 (2012).

\bibitem{amir2011huge} A. Amir, S. Borini, Y. Oreg, and Y. Imry, Phys. Rev.
Lett. {\bf 107}, 186407 (2011).

\bibitem{woltjer1993time} R. Woltjer, A. Hamada, and E. Takeda, Electron. Dev.,
IEEE Trans. {\bf 40}, 392 (1993).

\bibitem{sperl2003}
M. Sperl, Phys. Rev. E {\bf 68}, 031405 (2003).

\bibitem{angelani01} 
L. Angelani, R. Di Leonardo, G. Parisi, and G. Ruocco,
Phys. Rev. Lett. \textbf{87}, 055502 (2001).

\bibitem{supp} Supplementary Material.

\bibitem{chowdhury1984logarithmic}
D. Chowdhury and A. Mookerjee, J. of Phys. F: Met. Phys {\bf 14}, 245 (1984).

\bibitem{barabasi1999} A.-L. Barab{\'a}si, R. Albert and H. Jeong, Physica A
  {\bf 272}, 173 (1999).

\bibitem{havlin2002} 
S. Havlin and D. Ben-Avraham, Adv. Phys. {\bf 51}, 187 (2002).

\bibitem{zia1999} 
B. Schmittmann and R.K.P. Zia, Am. J. Phys. {\bf 67}, 1269 (1999).

\bibitem{REMM} The occurrence of the ratio $t/t_0$ in Eq.~(\ref{eq:mainres})
is similar to confined, subdiffusive continuous time random walk processes,
see S. Burov, R. Metzler, and E. Barkai, Proc. Natl. Acad. Sci. USA
\textbf{107}, 13228 (2010).

\bibitem{REM} Travel speeds in between cities is assumed to be much shorter
than forward waiting times.

\bibitem{REMMM} K. Briggs and C. Beck, Physica A \textbf{378}, 498 (2007).
 
\bibitem{oliveira2005human} J. G. Oliveira and A.-L. Barab{\'a}si, Nature
{\bf 437}, 1251 (2005).

\bibitem{barabasi2005origin} A.-L. Barab{\`a}si, Nature {\bf 435}, 207 (2005).

\bibitem{glarum} S. H. Glarum, J. Chem. Phys. \textbf{33}, 639 (1960).

\bibitem{dynkin} 
E. B. Dynkin, Izv. Akad. Nauk. SSSR Ser. Math. {\bf 19}, 247 (1955); Selected
Translations Math. Stat. Prob. {\bf 1}, 171 (1961).

\bibitem{godreche} C. Godr{\`e}che and J. M. Luck, J. Stat. Phys. \textbf{104},
489 (2001); E. Barkai and Y.-C. Cheng, J. Chem. Phys. \textbf{118}, 6167 (2003);
E. Barkai, Phys. Rev. Lett. \textbf{90}, 104101 (2003).

\bibitem{koren} T. Koren, M. A. Lomholt, A. V. Chechkin, J. Klafter, and R.
Metzler, Phys. Rev. Lett. \textbf{99}, 160602 (2007).

\bibitem{scher} E. W. Montroll and G. H. Weiss, J. Math. Phys. \textbf{6}, 167
(1965); H. Scher and E. W. Montroll, Phys. Rev. B \textbf{12}, 2455 (1975).

\bibitem{report} R. Metzler and J. Klafter, Phys. Rep. \textbf{339}, 1 (2000);
J. Phys. A \textbf{37}, R161 (2004).

\bibitem{davies2002} B. Davies, {\it Integral transforms and their
applications}, Springer (2002).

\bibitem{REM1} Our stochastic simulations are based on a forward jumping
process on a 1D lattice illustrated in Fig.~\ref{fig1}.

\bibitem{klaso} J. Klafter and I. M. Sokolov, First Steps in Random Walks
(Oxford, New York, NY, 2011).

\bibitem{REM2} For $\alpha<1$ the $t'$-dependence in $\psi_1(\tau_1|t')$
  hinders us from using Laplace-transforms which otherwise is useful for
  solving convolution problems. For $\alpha>1$ Eqs. (2) to (4) can be solved
  using this transform, yielding Eq. (\ref{eq:sublinear}).

\bibitem{pt} E. Barkai, Y. Garini, and R. Metzler, Physics Today {\bf 65}(8),
29 (2012).

\bibitem{bouchaud92} J.-P. Bouchaud, J. Phys. I France \textbf{2}, 1705 (1992).

\bibitem{bel07} G. Bel and E. Barkai, Phys. Rev. Lett. \textbf{94}, 240602 (2005).

\bibitem{js} J. H. P. Schulz, E. Barkai, and R. Metzler, Phys. Rev. Lett. \textbf{110},
020602 (2013).

\bibitem{limpert2001} E. Limpert, W.A. Stahel and M. Abbt, BioScience
{\bf 51}, 341 (2001).

\end{thebibliography}
\end{document}